\documentclass[aps,showpacs,showkeys,preprint]{revtex4}
\usepackage{mathrsfs}
\usepackage{amssymb}
\usepackage{amsmath}
\usepackage{bm}
\usepackage{graphics}
\usepackage{natbib}
\newcommand{\llangle}{\langle\!\langle}
\newcommand{\rrangle}{\rangle\!\rangle}

\newcommand{\sss}[1]{{\scriptscriptstyle #1}}
\begin{document}
\title{Contact conductance between graphene and quantum wires}
\author{Haidong Li}
\author{Yisong Zheng}\email[Corresponding author. Tel.: 011+86-431-849-9051;
Fax: 011+86-431-894-1554;
 Email address: ]{zys@mail.jlu.edu.cn (Y. Zheng)}
\address{Department of physics, Jilin University, Changchun 130023, China}
\date{\today}

\begin{abstract}
 The contact conductance between graphene and two
quantum wires which serve as the leads to connect graphene and
electron reservoirs is theoretically studied. Our investigation
indicates that the contact conductance depends sensitively on the
graphene-lead coupling configuration. When each quantum wire couples
solely to one carbon atom, the contact conductance vanishes at the
Dirac point if the two carbon atoms coupling to the two leads belong
to the same sublattice of graphene. We find that such a feature
arises from the chirality of the Dirac electron in graphene. Such a
chirality associated with conductance zero disappears when a quantum
wire couples to multiple carbon atoms. The general result irrelevant
to the coupling configuration is that the contact conductance decays
rapidly with the increase of the distance between the two leads. In
addition, in the weak graphene-lead coupling limit, when the
distance between the two leads is much larger than the size of the
graphene-lead contact areas and the incident electron energy is
close to the Dirac point, the contact conductance is proportional to
the square of the product of the two graphene-lead contact areas,
and inversely proportional to the square of the distance between the
two leads.
\end{abstract}

  \maketitle\bigskip

\section*{1. Introduction}

The fabrication of graphene, a single atomic layer of graphite, is
the first experimental realization of realistic two-dimensional
crystal \cite{refNovoselov1}. In the effective mass approximation,
valence electrons in such a carbon material obey the massless
relativistic Dirac equation, rather than the Schr\"{o}dinger
equation. Consequently, graphene presents many unusual electronic
transport properties, such as the half-integer quantum Hall effect
\cite{refzheng,refGusynin,refYb.Zhang,refNovoselov,refPeres}, the
nonzero conductivity minimum at vanishing carrier concentration
 \cite{refNovoselov,refM.I.Katsnelson1,refJ.Tworzydlo,refAKGeim,refFMiao,refYW.Tan,refK.Ziegler,refK.Ziegler2},
the subtle weak localization
\cite{refSuzuura,refS.V.Morozov,refA.F.Morpurgo,refKhveshchenko,refE.McCann,refXiaosongWu},
and the reflectionless transmission of the carrier through an
arbitrarily high barrier
\cite{refKrekora,refM.I.Katsnelson,refV.V.Cheianov}.
\par
The intriguing transport properties mentioned above are closely
associated with the scattering of a Dirac electron by impurities,
defects or gated barriers, {\it etc}. However, apart from these
scattering effects, the contact between graphene and the metallic
electrodes also influences the electronic transport spectrum to some
extent. In other words, the measured conductance(conductivity)
spectrum often includes a contact conductance, accompanying the
scattering conductance. From the experimental viewpoint, such a
contact conductance makes sense when graphene couples to the probes
of the scanning tunneling microscope (STM)
\cite{refKobayashi,refNiimi,refH.Amara}, or ultrathin gold or
tungsten wires; in particular, the multi-probe STM that was
developed quite recently \cite{refJ.Onoe,refK.Takami, refO.Guise,
refM.Ishikawa,refO.Kubo}, which can be used to explore the surface
structure of graphene. In these actual structures, the leads that
connect graphene and electron reservoirs are the quantum wires which
have only a few transverse modes to carry the current. This is in
contrast with the infinitely many transverse modes in graphene.
Thus, the electron will be reflected with certain probability when
it enters the leads from graphene. Besides, such quantum wires
couple locally to a finite number of the carbon atoms of graphene.
As a result, the electronic tunneling between the leads and graphene
depends sensitively on the coupling configuration at the contact. To
sum up these facts, it can be anticipated that the coupling between
graphene and a quantum wire will give a nontrivial contact
conductance.
\par
So far, relatively fewer works involve the contact conductance
between graphene and the leads of different kinds
\cite{refYMBlanter,refSchomerus}, in comparison with the scattering
conductance. In fact, many aspects concerning this issue deserve
further explorations. In the present work, we carry out a systematic
investigation on the contact conductance between graphene and
quantum wires which serve as the leads to conduct the current. There
are many factors that influence the contact conductance, for
example, the bandwidth and the band position of the quantum
wire(lead) relative to the Dirac point of graphene, the coupling
configuration between the leads and graphene, and the distance and
the relative orientation between the two leads. To obtain a
quantitative dependence of the contact conductance on these factors
it is desirable to analyse the observed conductance spectrum in
relevant experimental setups.
\par
The rest of this paper is organized as follows: In Sec.2, the
theoretical framework to formulate the contact conductance is
elucidated. In Sec.3, the numerical calculations on the contact
conductance are performed. Based on these numerical results, the
dependence of the contact conductance on the relevant parameters is
discussed. Finally, in Sec.4, we make some concluding remarks.

\par
\section*{2. Model and theory}
As illustrated in Fig.1, the electronic transport structure we
consider consists of a graphene monolayer coupling to two quantum
wires, which serve as two leads to conduct the current. We adopt a
semi-infinite linear lattice to describe the band structure of a
single transverse mode in each lead. Such a single mode lead couples
to the finite number
 of the carbon atoms in graphene. Note: Some structure
 parameters are explained in Fig.1.

In such a model the tight-binding Hamiltonian
 of an electron is composed of three parts,
\begin{equation}
   H=H_{C}+H_{G}+H_{T}. \label{1}
\end{equation}
The first term $H_{C}$ is the Hamiltonian of the two semi-infinite
linear lattices which model the two leads. It is given by
\begin{equation}
H_{C}=\sum_{j\leq-1}\varepsilon _{0}c_{j}^\dag c_{j}+\sum_{j<-1}(v_0
c_{j+1}^\dag c_{j}+{\mathrm {H.c.}})+\sum_{j\geq+1}\varepsilon
_{0}c_{j}^\dag c_{j} +\sum_{j\geq1}(v_0 c_{j+1}^\dag c_{j}+{\mathrm
{H.c.}}),\label{3}
\end{equation}
where $c^{\dagger}_{j}$ ($c_{j}$) is the electron creation
(annihilation) operator at the individual lattice points. For the
ideal leads, both the on-site energy and the hopping energy are
viewed as uniform parameters, denoted by $\varepsilon_0$ and $v_0$,
respectively. The second term $H_G$ is the tight-binding Hamiltonian
of the $\pi$ band electron in graphene. It takes a form as
\begin{equation}
H_{G}=t\sum_{\langle i,j\rangle}(d_{iA}^\dag d_{jB} +{\mathrm
{H.c.}}),\label{3}
\end{equation}
where $d^{\dagger}_{i\nu}$ $(d_{i\nu})$ with $\nu=A$ or B is the
electron creation(annihilation) operator associated with the local
atomic orbits in graphene. The notation $\langle i,j\rangle$ means
that the summation is restricted between the pairs of the nearest
neighbor carbon atoms. $t$ is the corresponding hopping energy. We
choose the Dirac point as the energy reference point, hence the
on-site energy of all lattice points in graphene is zero. Moreover,
in what follows we use the units $\hbar=t=a=1$. The last term
$H_{T}$ in the total Hamiltonian describes the electron tunneling
between the leads and graphene. It is given by
\begin{equation}
H_{T} =\sum_{i\nu}(v_{\sss{L}i\nu } c^\dag_{-1}d_{i\nu}+{\mathrm
{H.c.}})+\sum_{j\nu}(v_{\sss{R} j\nu} c^\dag_{1}d_{j\nu} +{\mathrm
{H.c.}}),
\end{equation}
where $v_{\sss{L}i\nu}$ and $v_{\sss{R}j\nu}$ denote the coupling
strength between two leads and the individual carbon atoms in
graphene. Note that we use L and R to denote the two leads
respectively. As illustrated in Fig.1, only finite carbon atoms
around the tip of each lead couple to the lead effectively.
\par

To study the electronic transport properties, we need to calculate
the linear conductance between the two leads. Based on the
Landauer-B\"{u}ttiker theory \cite{refLandauer,refM.Buttiker}, we
can write the linear conductance formula as
\begin{equation}
\mathcal
G(E)=\frac{2e^{2}}{h}\mathrm{Tr}[\bm{\Gamma}^\sss{L}(E)\bm{G}^r(E)\bm{\Gamma}^\sss{R}(E)\bm{G}^a(E)],\label{CF}
\end{equation}
where $E$ is the energy of the incident electron from one lead. The
matrices of the retarded and advanced Green functions satisfy a
relationship $\bm{G}^r(E)^\dag=\bm{G}^a(E)$. The matrix element is
given by $[\bm{G}^r]_{i\nu_{1}j\nu_{2}}=\int^{\infty}_{-\infty}
G_{i\nu_{1}j\nu_{2}}^{r}(t)e^{iEt}dt$ and
$G_{i\nu_{1}j\nu_{2}}^r(t)=-i\theta(t)
\langle\{d_{i\nu_{1}}(t),d_{j\nu_{2}}^\dag\}\rangle$, just following
the standard definition of a retarded Green function. We will often
use an alternative notation $\llangle A|B \rrangle^r$ to denote the
retarded Green function in Fourier space for convenience, e.g.
$[\bm{G}^r]_{i\nu_{1}j\nu_{2}}=\llangle
d_{i\nu_1}|d^{\dagger}_{j\nu_2}\rrangle^r$. In Eq.(\ref{CF}) two
other matrices are defined as $[\bm{\Gamma}^{\sss{L}}]_{i\nu,
i'\nu'}=-2{v_{Li\nu}v_{Li'\nu'}^{*}}\text{Im}\llangle
c_{-1}|c^{\dagger}_{-1} \rrangle^r_{0}$ and
$[\bm{\Gamma}^{\sss{R}}]_{j\nu,
j'\nu'}=-2{v_{Rj\nu}v_{Rj'\nu'}^{*}}\text{Im}\llangle
c_{1}|c^{\dagger}_{1} \rrangle^r_{0}$, where the subscript ``0"
denotes that the corresponding Green function belongs to an
individual lead, completely isolated from graphene. Besides, as we
have done, we will often drop the energy dependence of these
matrices to keep expressions compact.
\par
To solve the Green functions in the conductance formula, we need to
convert the Hamiltonian $H_G$ into the eigen-representation of the
$\pi$ band electron of graphene. Doing so, we utilize the following
transformations
\begin{equation}
d_{i\nu}=\frac{1}{\sqrt{N}}\underset{k}{\sum}e^{i\bm{k}\cdot\bm{R}_{i\nu}}c_{k\nu},\label{transmission}
\end{equation}
and
\begin{equation}
c_{kA}=\underset{s=\pm}\sum\frac{s t_{k}}{\sqrt{2}\mid
t_{k}\mid}\alpha_{ks},\; \;
c_{kB}=\underset{s=\pm}\sum\frac{1}{\sqrt{2}}\alpha_{ks},\label{transmission}
\end{equation}
where $N$ is the total number of unit cells in graphene;
$t_k=\sum_{l=1}^3t\exp(i\bm{k}\bm{\tau}_l)$ with $\bm{k}$ being the
electron wave vector measured from the center of the Brillouin zone
of graphene. The creation (annihilation) operator
$c^\dag_{k\nu}$($c_{k\nu}$) is associated with electron Bloch state
in one sublattice of graphene. And the notation $s=+$ or $-$,
denotes the conduction or valence bands of graphene, respectively.
$\alpha^{\dagger}_{ks} (\alpha_{ks})$ is the creation (annihilation)
operator of the eigen-states of the valence electron in graphene.
Thus, the Hamiltonian of graphene becomes diagonal
\begin{equation}
H_{G}=\underset{k,s}\sum
\varepsilon_{ks}\alpha^{\dag}_{ks}\alpha_{ks},\label{transmission}
\end{equation}
where $\varepsilon_{ks}=s|t_k|$. Accompanying such a representation
transformation, the tunneling Hamiltonian $H_T$ can be rewritten as
\begin{equation}
H_{T}=\sum_{ks}(V_{Lks}c^{\dagger}_{-1}\alpha_{ks}+{\mathrm
{H.c.}})+\sum_{ks}(V_{Rks}c^{\dagger}_{1}\alpha_{ks}+{\mathrm
{H.c.}}),\label{3}
\end{equation}
with
\begin{equation}
V_{\mu ks}=\underset{i}\sum\frac{v_{\mu
i\sss{A}}}{\sqrt{2N}}e^{i\bm{k} \bm{R}_{iA}}\frac{s t_{k}}{\mid
t_{k}\mid}+\underset{i}\sum\frac{v_{\mu
i\sss{B}}}{\sqrt{2N}}e^{i\bm{k}\bm{R}_{iB}}, \;\; \mu=L,
R.\label{vmuks}
\end{equation}
Meanwhile, the conductance expression changes into
\begin{equation} \mathcal
G(E)=\frac{2e^{2}}{h}\left|\sum_{ks}\sum_{k's'}V_{Lks}\llangle
\alpha_{ks}|\alpha^{\dagger}_{k's'}\rrangle^rV^{\ast}_{Rk's'}\right
|^24\mathrm{Im}\llangle
c_{-1}|c^{\dagger}_{-1}\rrangle_{0}^r\mathrm{Im}\llangle
c_{1}|c^{\dagger}_{1}\rrangle_{0}^r.\label{3}
\end{equation}
The retarded Green function satisfies the following equation of
motion
\begin{eqnarray}
\llangle\alpha_{ks}|\alpha^\dag_{k's'}\rrangle^r&=&g_{ks}\delta_{ks,k's'}
+g_{k's'}V_{Lk's'}g_{ks}V^{\ast}_{Lks}\llangle
c_{-1}|c^{\dagger}_{-1} \rrangle^r
+g_{k's'}V_{Lk's'}g_{ks}V^{\ast}_{Rks}\llangle
c_{1}|c^{\dagger}_{-1} \rrangle^r\notag\\& &
+g_{k's'}V_{Rk's'}g_{ks}V^{\ast}_{Lks}\llangle
c_{-1}|c^{\dagger}_{1} \rrangle^r
+g_{k's'}V_{Rk's'}g_{ks}V^{\ast}_{Rks}\llangle c_{1}|c^{\dagger}_{1}
\rrangle^r,\label{transmission}
\end{eqnarray}
where $g_{ks}=(E^+-\varepsilon_{ks})^{-1}$ and $E^+=E+i\eta$ with
$\eta$ being a positive infinitesimal. Here, we can see that the
Green functions $\llangle c_{\lambda}|c^{\dagger}_{\lambda'}
\rrangle^r$ $(\lambda,\lambda'=\pm1)$ are the relevant quantities to
calculate the linear conductance, which are exactly solvable. Taking
$\llangle c_{-1}|c^{\dagger}_{-1} \rrangle^r$ as an example, the set
of the equation-of-motion with regard to this Green function
consists of
\begin{equation}
(E^+-\varepsilon_{0}-\xi_{LL})\llangle
c_{-1}|c^{\dagger}_{-1}\rrangle^r=1+v_{0}\llangle
c_{-2}|c^{\dagger}_{-1} \rrangle^r+\xi_{LR}\llangle
c_{1}|c^{\dagger}_{-1} \rrangle^r,\label{3}
\end{equation}
and
\begin{equation}
(E^+-\varepsilon_{0})\llangle
c_{-(j+1)}|c^{\dagger}_{-1}\rrangle^r=v_{0}\llangle
c_{-(j+2)}|c^{\dagger}_{-1} \rrangle^r+v_{0}\llangle
c_{-j}|c^{\dagger}_{-1} \rrangle^r \;\;\; \text{for}\;\;\; \
j=1,2,... .\label{3}
\end{equation}
In the above equations
\begin{equation}
\xi_{\mu\mu'}=\underset{ks}\sum V_{\mu ks}g_{ks}V^{\ast}_{\mu'
ks},\;\;\;\mu, \;\mu'=L\; \text{or}\; R.\label{xi}
\end{equation}
By virtue of the property of a tri-diagonal matrix \cite{refKuo}, we
can solve analytically the above set of equations, which yields
\begin{equation}
\llangle
c_{-1}|c^{\dagger}_{-1}\rrangle^r=\frac{\omega-\xi_{RR}}{(\omega-\xi_{LL})(\omega-\xi_{RR})-\xi_{LR}\xi_{RL}},\label{3}
\end{equation}
with
\begin{equation}
\omega=E-\varepsilon_{0}-\beta^*,
\end{equation}
and
\begin{equation}
\beta^{*}=\frac{E-\varepsilon_{0}-i\sqrt{4v^2_{0}-(E-\varepsilon_{0})^2}}{2}.
\end{equation}
In similar manner, we can obtain the analytical forms of Green
functions $\llangle c_{-1}|c^{\dagger}_{1}\rrangle^r$, $\llangle
c_{1}|c^{\dagger}_{-1}\rrangle^r$ and $\llangle
c_{1}|c^{\dagger}_{1}\rrangle^r$. They are given by
\begin{equation}
\llangle
c_{-1}|c^{\dagger}_{1}\rrangle^r=\frac{\xi_{LR}}{(\omega-\xi_{LL})(\omega-\xi_{RR})-\xi_{LR}\xi_{RL}},
\end{equation}
\begin{equation}
\llangle
c_{1}|c^{\dagger}_{-1}\rrangle^r=\frac{\xi_{RL}}{(\omega-\xi_{LL})(\omega-\xi_{RR})-\xi_{LR}\xi_{RL}},
\end{equation}
and
\begin{equation}
\llangle
c_{1}|c^{\dagger}_{1}\rrangle^r=\frac{\omega-\xi_{LL}}{(\omega-\xi_{LL})(\omega-\xi_{RR})-\xi_{LR}\xi_{RL}}.\label{3}
\end{equation}
Substituting these results into Eq.(\ref{transmission}), we can then
get the explicit form of the Green function
$\llangle\alpha_{ks}|\alpha^\dag_{k's'}\rrangle^r$. Subsequently,
after some derivation, we obtain an analytical expression for the
contact conductance. That is
\begin{equation}
\mathcal G(E)=\frac{2e^{2}}{h}T(E),\label{3}
\end{equation}
\begin{eqnarray}
T(E)=|\rho_0(E)\tilde{t}(E)|^2,\label{te}
\end{eqnarray}
with
\begin{equation}
\tilde{t}(E)=\frac{\omega^2\xi_{LR}}{(\omega-\xi_{LL})(\omega-\xi_{RR})-\xi_{LR}\xi_{RL}}
\label{trans},
\end{equation}
and
\begin{equation}
\rho_0(E)=-2\text{Im}\llangle
c_{-1}|c^{\dagger}_{-1}\rrangle_{0}^r=\frac{\sqrt{4v^{2}_{0}-(E-\varepsilon_{0})^{2}}}{v^{2}_{0}}.
\label{dos}
\end{equation}
So far we have obtained a conductance expression in terms of the
self-energy terms $\xi_{\mu \mu'}$. We will derive their the
analytical forms in the Appendix. To calculate the conductance
spectrum, the two parameters $v_0$ and $\varepsilon_0$ ought to be
taken the appropriate values to guarantee the energy band of the
leads and the linear region of the $\pi$ band of graphene overlaps
each other. Besides, the incident electron energy should also be
restricted in the linear region of graphene band. Now we consider an
extreme case that the graphene-lead coupling is far much smaller
than the bandwidth of the leads, i.e. $v\ll v_0$. From the above
conductance expression, we can infer that the conductance formula in
such a weak coupling limit reduces to
\begin{equation}
\mathcal G(E)=\frac{2e^{2}}{h}|\rho_0(E)\xi_\sss{LR}|^2
.\label{wideband}
\end{equation}

\section*{3. Results and discussion \label{result}}
After formulating the linear conductance, we are now in a position
to perform the numerical calculation for the conductance spectrum,
from which we can investigate the electronic transport properties
dominated by the contacts between graphene and two quantum wires.
\par
First of all, we consider the simplest case that only one carbon
atom in graphene couples to each lead(Hereafter we call the tip of a
lead a probe). For the numerical calculation, we fix the first
probe(L) to couple to the $A$ atom at origin, and shift the second
probe around. In Fig.\ref{a1a1}, we show the calculated contact
conductance spectrum ($\mathcal G(E)$ versus $E$) for the second
probe(R) stopping at a specific A atom. In such a case we have
$v_{\sss{L}i\nu}=v\delta_{i0}\delta_{\nu\sss{A}}$ and
$v_{\sss{R}j\nu}=v\delta_{jm}\delta_{\nu\sss{A}}$. We refer to such
a configuration between graphene and two leads as A-A coupling. The
most striking feature shown in this figure is that the contact
conductance vanishes when the incident electron energy is aligned
with the Dirac point of graphene. From the analytical results given
in the Appendix, we can readily obtain the self-energy terms
$\xi_{\mu\mu'}$ for the case of the simple A-A coupling
\begin{equation}
\xi_{\sss{LL}}=\xi_{\sss{RR}}=\frac{\sqrt{3}v^2E}{2\pi}\int_0^{q_c}
\frac{qdq}{{E^{+}}^2-(\gamma q)^2},
\end{equation}
\begin{equation}
\xi_{\sss{LR}}=\frac{\sqrt{3}v^2E\cos(\frac{4\pi R_x}{3})}
{2\pi}\int_0^{q_c}\frac{J_{0}(qR)qdq}{{E^{+}}^2-(\gamma
q)^2}.\label{xilraa}
\end{equation}
Owing to the chirality of the $\pi$ band electron in graphene, these
self-energies vanish at Dirac point($E=0$). Then, this leads to the
zero point of the contact conductance as shown in Fig.\ref{a1a1}.
Therefore, we can attribute the zero contact conductance at the
Dirac point to the chirality of the Dirac electron. As shown in
Fig.\ref{a1a1}(a), we can shift the position of the band bottom of
the leads relative to the Dirac point by varying the parameter
$\varepsilon_0$. As a result, the conductance spectrum becomes
asymmetric with respect to the Dirac point. In addition,
Fig.\ref{a1a1}(c) shows that when the band of the leads is widened
by increasing the parameter $v_0$, the conductance becomes notably
smaller. Such a dependence of the contact conductance spectrum on
the band structure of the leads can be readily understood by
analyzing the local density of states of the electron at the tip of
a lead, which appears in the conductance formula, see Eqs.(\ref{3}),
(\ref{te}) and (\ref{wideband}). From Eq.(\ref{dos}) we can see that
such a local density of states decreases as the bandwidth of a lead
gets larger. Besides, when the parameter $\varepsilon_0$ deviates
from the Dirac point, the local density of states becomes asymmetric
relative to the Dirac point. The variation of the conductance
spectra with the parameters $v_0 $ and $\varepsilon_0$ shown in
Fig.\ref{a1a1}(a) and (b) just reflects these features of the local
density of states. In Fig.\ref{a1a1}(c) and (d) we show the contact
conductance spectrum for the so-called A-B coupling which means that
the second probe couples solely to a $B$ atom at a specific
position. In such a case the band structure of the leads influences
the conductance spectrum in the same way as the case of A-A
coupling, namely, the conductance spectrum becomes asymmetric with
the shift of $\varepsilon_0$ and decreases globally with the
increase of $v_0$. However, for the case of A-B coupling a
conductance peak occurs at Dirac point in place of the conductance
zero in A-A coupling. This is due to that the self-energy
$\xi_{\sss{LR}}$ for A-B coupling takes a different form. It is
given by
\begin{equation}
\xi_{\sss{LR}}=\frac{\sqrt{3}\gamma\upsilon^{2}\sin(\frac{4\pi
R_x}{3}+\theta)}
{2\pi}\int_0^{q_c}\frac{J_1(qR)q^2dq}{{E^{+}}^2-(\gamma
q)^2}.\label{xilrab}
\end{equation}
By a simple derivation, we can further deduce that for the A-B
coupling the self energies at the Dirac point are
$\xi_{\sss{LL}}=\xi_{\sss{RR}}=0$ and
\begin{equation}
\xi_\sss{LR}=\frac{\sqrt{3}\upsilon^{2}\sin({4\pi R_x\over
3}+\theta)}{2\pi\gamma R}[1-J_0(q_cR)].
\end{equation}
Then the contact conductance at the Dirac point takes a simple form
as
\begin{equation}
\mathcal
{G}(E=0)=(\frac{2\upsilon_{0}\xi_\sss{LR}}{\upsilon^2_{0}+\xi^2_\sss{LR}})^2.
\label{cond_e0}
\end{equation}
\par
With the help of these self-energy terms we can discuss the
dependence of the contact conductance on the distance between the
two probes. In fact, from Eqs.(\ref{xilrab})-(\ref{cond_e0}) we can
infer that initially with the increase of the distance between the
two probes, the contact conductance near the Dirac point will decay
rapidly. However, when the distance between the two probes becomes
sufficiently large, the contact conductance tends to be inversely
proportional to $R^2$. In Fig.\ref{c_r}(a) we plot the contact
conductance as a function of $R$ by letting the second probe to move
away from the origin (the position of the first probe) along the y
axis. We can see that the contact conductance decreases drastically
when $R$ increases within several times the lattice constant. Such a
rapid decay has little to do with the variation of the incident
electron energy.
\par
The explicit expression of the contact conductance at the Dirac
point shown by Eq.(\ref{cond_e0}) indicates a conductance peak
occurs when $v_{0}=\xi_\sss{LR}$. This implies that the resonance
will occur when the distance between the two probes takes an
appropriate value, which depends on the bandwidths of the leads and
their coupling strengths to graphene. To study such a resonance in
some details, we calculate the conductance as a function of the
strength of the graphene-lead coupling, as shown in
Fig.\ref{resonance}. From Fig.\ref{resonance}(a) we can see that a
stronger graphene-lead coupling is required to observe resonance
when the distance between the two probes gets larger. The calculated
results shown in Fig.\ref{resonance}(b) indicate that the resonant
conductance peak is notably suppressed when the incident electron
energy deviates from the Dirac point.
\par
Now we turn to discuss the orientation dependence of the linear
conductance when the second probe shifts around the origin where the
first probe is located. For simplicity, we only consider the weak
graphene-lead coupling limit, where the conductance expression takes
a relatively simple form, given by Eq.(\ref{wideband}). From
Eqs.(\ref{xilraa}) and (\ref{xilrab}) we can see that the factors
$\cos({4\pi R_x\over 3})$ for A-A coupling and $\sin({4\pi R_x\over
3}+\theta)$ for A-B coupling determine the orientation dependence of
the linear conductance. Our derivation in the Appendix indicates
that such orientation factors originate from quantum interference
between the K and K' valleys. Consequently, when the length of
$\bm{R}$(the distance between the two probes) is fixed, the maximum
of the conductance appears when $\bm{R}$ is along $y$ direction.
Furthermore, considering the symmetry of the honeycomb lattice of
graphene, when the second probe moves along a circle around the
first probe, the conductance maximum will appear at the
$\pm\bm{\tau}_l$ directions. In addition, we can readily infer that
for A-A coupling the conductance pattern formed by shifting the
second probe has reflection symmetry with respect to the x axis. In
other words, when the second probe is located at the two symmetric A
atoms relative to the x axis, the conductance gives the same value.
\par
The above coupling manner that a quantum wire couples to a single
carbon atom in graphene can only model the extreme situation that
the probe aims at a specific carbon atom, but the coupling strength
between the probe and the adjacent carbon atoms is negligibly small.
To mimic the actual graphene-lead coupling configuration, we
introduce the following Gaussian-type graphene-lead coupling
\begin{equation}
v_{\mu i\nu}=v\exp[-{(\bm{r}-\bm{R}_{i\nu})^2\over d_0^2}],\;\;
\mu=L,\; R \;\; \text{and}\;\;  \nu=A, \; B, \label{gauss}
\end{equation}
where $\bm{r}$ denotes the arbitrary position of probe $\mu$, and
$\bm{R}_{i\nu}$ is the lattice vector of a carbon atom in the
vicinity of the probe. $d_0$ is a decaying factor to determine the
region in which the carbon atoms couple effectively to the probe.
Obviously, when $d_0$ is far smaller than the lattice constant, such
a Gaussian-type coupling changes into the single atom coupling
configuration discussed above. On the other hand, when $d_0$ is
comparable to or larger than the lattice constant, a probe will
couple to multiple carbon atoms around it.
\par

At first, we check whether the contact conductance vanishes at the
Dirac point when the two probes are positioned at two carbon atoms
of the same kind. As shown in Fig.\ref{c_r}(b), we can see that such
a chirality associated feature no longer exists when the effective
coupling region gets larger. This is readily understood since in the
Gaussian-type coupling configuration with a decay factor $d_0$
comparable to the lattice constant of graphene, the conductance is
the averaging result of A-A and A-B couplings. Besides, the result
shown in Fig.\ref{c_r}(b) indicates that the rapid decay of the
contact conductance with the increase of the probe interval still
holds in such a Gaussian-type graphene-lead coupling. In
Fig.\ref{resonance}(c) we show the resonant contact conductance by
changing the strength of the graphene-lead coupling in Gaussian
type. We can infer that the resonant conductance peak is influenced
by the decay factor $d_0$ intricately. Initially when we increase
the decay factor($d_0=0.4$ to 0.8) there is a notable decrease in
the resonant peak. But when we increase $d_0$ further, the
conductance peak turns to get larger. Such a complicated dependence
of the conductance on the effective coupling size arises from the
quantum interference between the different electron transmission
paths. When an electron travels from the first probe to the second
one, electronic tunnelings between graphene and the two probes can
be realized via two carbon atoms of either the same kind or the
distinct kind, which corresponds to the different electron
transmission paths. These two distinct electron transmission paths
result in destructive interference. At $d_0=0.8$, besides the
central B atom, three nearest neighbor A atoms begin to couple
effectively to the second probe. As a result, the destructive
quantum interference diminishes the resonant peak. When $d_0=1.2$,
six next-nearest neighbor B atoms enter the effective region to
couple to the second probe, which compensate for the negative effect
of the three nearest neighbor A atoms. As a result, there is a
notable increase in the resonant peak.
\par
With such a Gaussian-type coupling manner we can plot the
two-dimensional conductance pattern formed by shifting the second
probe around the first one. In Fig.\ref{patternd02} we plot such a
conductance pattern for a relatively small decay factor in the
Gaussian-type coupling function($d_0=0.2$). This case is in analogy
to the single atom coupling configuration discussed above. From this
figure we can see that the conductance pattern exhibits the $C_3$
group symmetry when the second probe rotates around the first probe.
And such a rotation symmetry is independent of the incident electron
energy. The strongest conductance appears when the second probe is
located at the next-nearest neighbor B atoms, rather than the three
nearest neighbor B atoms around the origin (Except for the
conductance maximum at the origin when the incident electron energy
is far away from the Dirac point, as shown in
Fig.\ref{patternd02}(b)). According to the discussion on the
relation between the resonant conductance peak and the probe
interval, the strongest conductance can appear at different
positions if we adjust the coupling strength between the probes and
graphene. In Fig.\ref{patternd1} we plot the conductance pattern
corresponding to a relatively large decay factor($d_0=1.0$). We can
see that the $C_3$ group symmetry remains in this conductance
pattern. And conductance pattern does not vary notably with the
variation of the incident electron energy. Unlike the case of
$d_0=0.2$ where the conductance maxima always occur when the second
probe points at an individual carbon atom, some discrete islands
with different shapes form in Fig.\ref{patternd1}, labeling the
regions with the relatively large conductance. As shown in
Fig.\ref{patternring}, if we move the first probe to the center of a
hexagon of the graphene lattice, the conductance pattern formed by
shifting the second probe around has the $C_6$ group symmetry
instead of the $C_3$ group symmetry in the previous cases shown in
Fig.\ref{patternd02} and Fig.\ref{patternd1}. However, in such a
case the conductance is far smaller than those in the previous
cases. This is because that the first probe does not point at any
atom carbon. Thus, according to the Gaussian-type coupling function,
the coupling strength between the carbon atoms in graphene and this
lead diminishes notably.
\par
Finally, based on the above discussion, we can work out an
approximate expression for the contact conductance, which is
convenient for a rapid estimation of the contact conductance for
some experimental setups if the following conditions are satisfied.
First, if the distance between the two probes, symbolized by
$\bar{\bm{R}}$, is much larger than the size of the effective
contact area around each probe, thus to calculate the self-energies
from the formulae given in the Appendix, we can replace the exact
distance between two individual carbon atoms around two probes by
$\bar{\bm{R}}$ approximately. Second, when the incident electron
energy is limited in the vicinity of the Dirac point, we can roughly
view $E\approx 0$. Thus, according to the formulae shown in the
Appendix, the dominant contribution to the self-energy arises from
$F_{\sss{AB}}(\bar{\bm{R}})$ and $F_{\sss{BA}}(\bar{\bm{R}})$.
Consequently, we arrive at an approximate expression about the
contact conductance in the weak coupling limit, which is given by
\begin{equation}
\mathcal{G}={2e^2\over h}{n_\sss{L}^2n_\sss{R}^2v^4\over
4v_0^2}|F_{\sss{BA}}|^2,
\end{equation}
where $n_\mu$ denotes the number of the same kind in graphene which
couple effectively to probe $\mu$. Obviously, such a quantity is
proportional to the contact area between graphene and the probe. If
the probe interval $\bar{R}$ is much larger than the lattice
constance of graphene, we will have
\begin{equation}
F_{\sss{BA}}={\sqrt{3}\sin({4\pi\bar{R}_x\over 3}+\bar{\theta})\over
2\pi\gamma \bar{R}},
\end{equation}
where $\bar{\theta}$ is the argument of $\bar{\bm{R}}$. Therefore,
we can conclude that the contact conductance is proportional to the
square of the two contact areas between graphene and the two probes,
and inversely proportional to the square of the probe interval, if
the conditions given above are satisfied.

\section*{4. Conclusions\label{Conclusions}}
In this work we have systematically studied the contact conductance
between graphene and two quantum wires which serve as the leads to
connect electron reservoirs and graphene. The general conclusion we
have obtained is that the contact conductance decays rapidly with
the increase of the distance between the two leads. When each
quantum wire couples to only one carbon atom in graphene, the
contact conductance vanishes at the Dirac point if the two carbon
atoms coupling respectively to the two leads belong to the same
sublattice of graphene. And this conductance zero arises from the
chirality of the Dirac electron in graphene. Also at the Dirac
point, if two quantum wires couple to two carbon atoms of distinct
kinds, a resonant path can be formed by adjusting the strength of
the graphene-lead coupling. The inter-valley quantum interference
causes the orientation dependence of the contact conductance. In the
weak graphene-lead coupling limit, the carbon-carbon bond
directions(i.e. $\bm{\tau}_l$ directions) are the optimal directions
to form the maximal contact conductance. In a more realistic
situation, each quantum wire may couple effectively to multiple
carbon atoms around it. To mimic such a situation, we introduce
Gaussian-type graphene-lead coupling, by which we have worked out
the two-dimensional conductance pattern formed by moving the second
probe around the first one. We find that the conductance pattern
does not vary sensitively with the incident electron energy.
However, the symmetry of the conductance pattern changes from $C_3$
group to $C_6$ group when the first probe shifts from a carbon atom
to the center of a hexagon of the graphene lattice. Finally, we
obtain an approximative expression about the contact conductance
when the probe interval is sufficiently large and the incident
electron energy is in the vicinity of the Dirac point. We find that
in such a case the contact conductance is proportional to the square
of the two contact areas between graphene and the probes, and
inversely proportional to the square of the probe interval.
\par

\section*{Acknowledgements}
  This work was financially supported by the National Natural Science
Foundation of China under Grant NNSFC10774055.

\section*{Appendix}

Now we derive the explicit expression about the self-energy terms
$\xi_{\mu\mu'}$
 in terms of the structure parameters. Substituting Eq.(\ref{vmuks}) into Eq.(\ref{xi})
 we have
 \begin{equation}
\xi_{\mu\mu'}=\underset{ks}\sum\frac{V_{\mu ks}V^{*}_{\mu'
ks}}{E^+-\varepsilon_{ks}}=\underset{ij}\sum v_{\mu i \nu}
F_{\nu\nu'}(\bm{R}_{i\nu}-\bm{R}_{j\nu'})v^*_{\mu' j \nu'}.
\end{equation}
In the above we have introduced four auxiliary functions
$F_{\nu\nu'}(\bm{R})$ which can be analytically treated by invoking
the linear dispersion relation of graphene around the Dirac point.
\begin{eqnarray}
&&F_{\sss{AA}}(\bm{R})=\frac{1}
{2N}\underset{ks}\sum\frac{e^{i\bm{k}\bm{R}}}{E^+-\varepsilon_{ks}}
=\frac{\sqrt{3}Ecos(\frac{4\pi R_x}{3})}
{2\pi}\int_0^{q_c}\frac{J_{0}(qR)q\,\mathrm{d}q}{{E^{+}}^2-(\gamma
q)^2},\label{faa}\\
&&F_{\sss{BB}}(\bm{R})=F_{\sss{AA}}(\bm{R}),\label{faa}
\end{eqnarray}
\begin{equation}
F_{\sss{AB}}(\bm{R})=\frac{1}{2N}\underset{ks}\sum\frac{e^{i\bm{k}
\bm{R}}}{E^+-\varepsilon_{ks}} \frac{s t_{k}}{\mid t_{k}\mid}
=\frac{\sqrt{3}\gamma M_{+}}
{2\pi}\int_0^{q_c}\frac{J_{1}(qR)q^2\,\mathrm{d}q}{{E^{+}}^2-(\gamma
q)^2},\label{fab}
\end{equation}
and
\begin{equation}
F_{\sss{BA}}(\bm{R})=\frac{1}{2N}\underset{ks}\sum\frac{e^{i\bm{k}
\bm{R}}}{E^+-\varepsilon_{ks}} \frac{s t^{*}_{k}}{\mid t_{k}\mid}
=\frac{\sqrt{3}\gamma M_-}
{2\pi}\int_0^{q_c}\frac{J_{1}(qR)q^2\,\mathrm{d}q}{{E^{+}}^2-(\gamma
q)^2},\label{fba}
\end{equation}
where $\gamma=\sqrt{3}at/2$ is the so-called Fermi velocity and
\begin{equation}
M_{\pm}=\sin(\frac{4\pi R_x}{3}\pm\theta).
\end{equation}
It should be noticed that in Eq.(\ref{faa}) the incident electron
energy $E$ appears in the analytical result of the functions
$F_{\sss{AA}}(\bm{R})$ and $F_{\sss{BB}}(\bm{R})$ as a prefactor. It
arises from the chirality of the Dirac electron in graphene. This
result implies that these two functions are equal to zero when the
incident electron energy is aligned with the Dirac point. Besides,
our derivation indicates that the quantum interference between K and
K' valleys in the band structure of graphene is responsible for the
dependence of the four functions given above, hence the contact
conductance, on the relative orientation between the two probes.

\clearpage \clearpage
\begin{figure}[htb]
\begin{center}
\scalebox{0.5}{\includegraphics{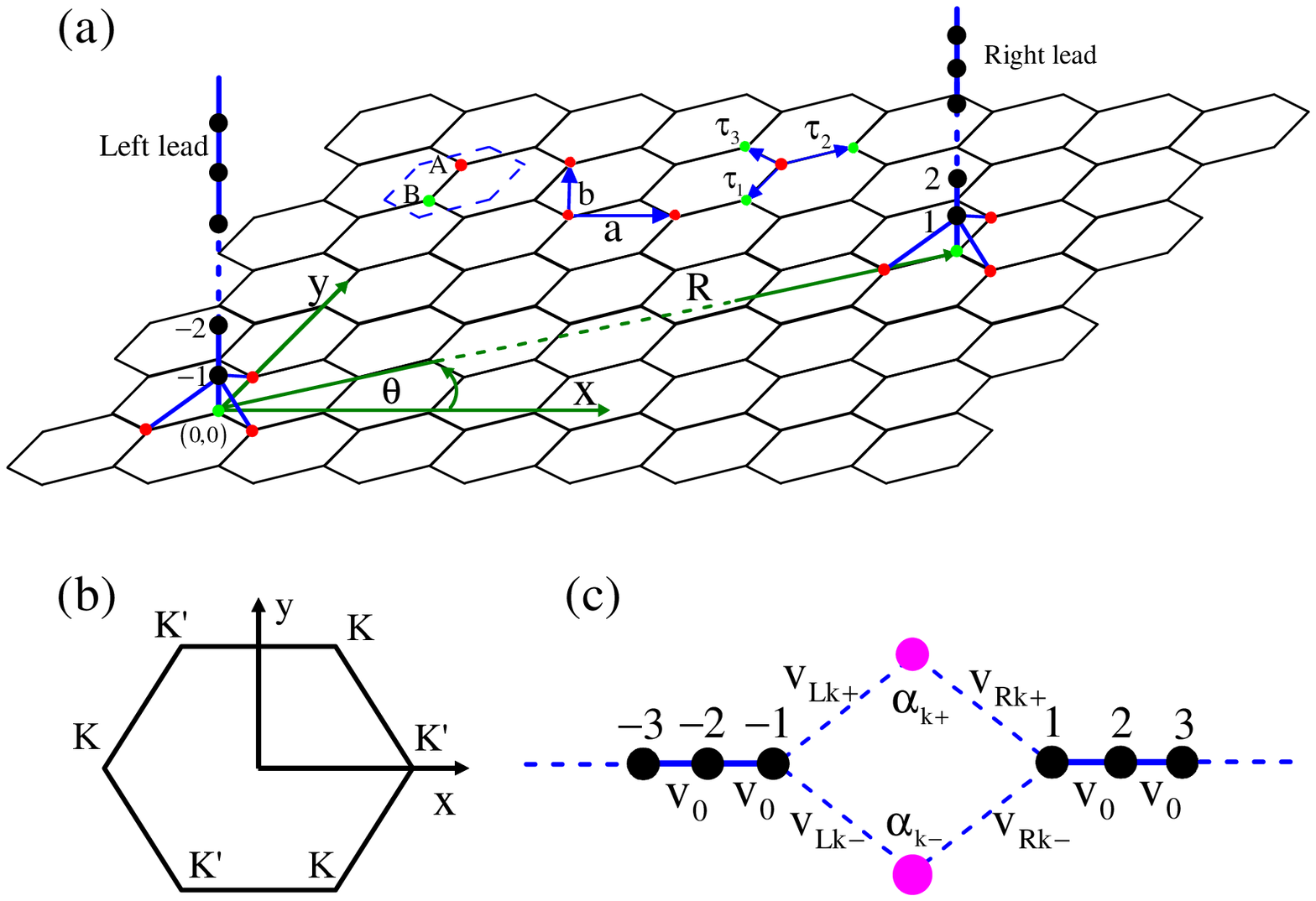}}\caption{(Color online)
(a) Schematic of the lattice structure of graphene and the reference
frame. A hexagonal unit cell represented by a dashed line contains
two carbon atoms denoted by A and B. Three vectors directed from a A
site to nearest neighbor B sites are given by
$\bm{\tau}_{1}=a[0,-1/\sqrt{3}]$,
$\bm{\tau}_{2}=a[1/2,1/(2\sqrt{3})]$ and
$\bm{\tau}_{3}=a[-1/2,1/(2\sqrt{3})]$ with $a$ being the lattice
constant. $\bm{R}$ is the distance between the centers of the two
leads. (b) The first Brillouin zone of graphene. The two
unequivalent vertices of the hexagon are called $K$ and $K^{'}$
points. (c) An equivalent plot of the electronic transport structure
shown in (a), but in (c) graphene is represented by two kinds of
eigen-states of the $\pi$ band electron. \label{structure}}
\end{center}
\end{figure}
\begin{figure}
\begin{center}
\scalebox{0.5}{\includegraphics{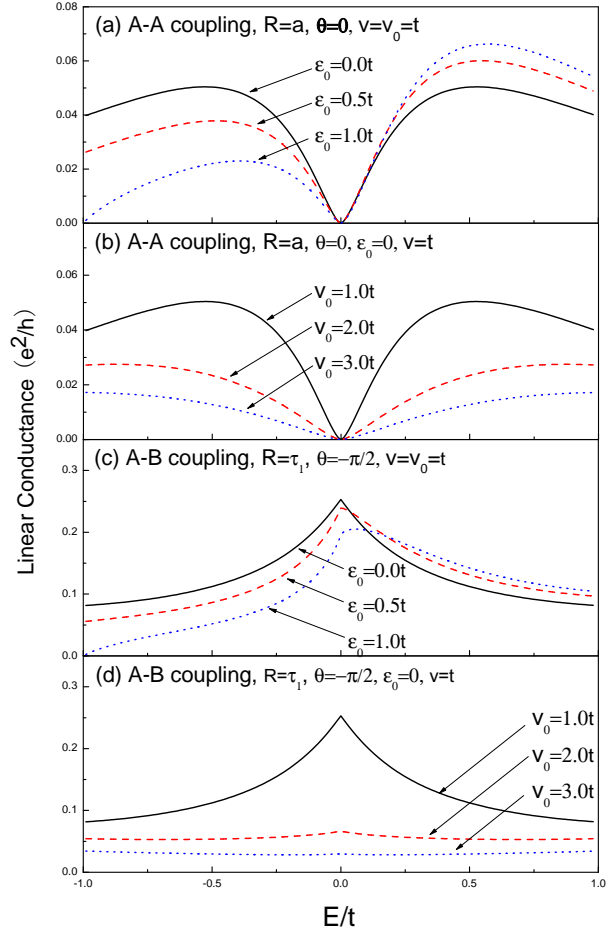}}\caption{(Color online)
The linear conductance as a function of the incident electron energy
for A-A coupling shown in (a) and (b), and A-B coupling shown in (c)
and (d). In (a) and (c) the parameter $\varepsilon_0$ takes several
typical values which shift the energy band of the leads relative to
the Dirac point of graphene. In (b) and (d) the parameter $v_0$
takes several typical values which correspond to different
bandwidthes of the leads. \label{a1a1}}
\end{center}
\end{figure}
\begin{figure}
\begin{center}
\scalebox{0.5}{\includegraphics{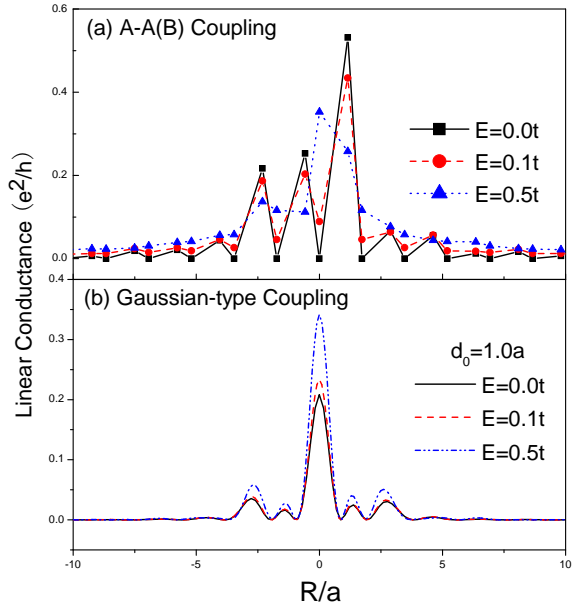}}\caption{(Color online)
The linear conductance as a function of R, the interval between the
two probes. The first probe is fixed at the origin. And the second
probe departs from the origin along the y direction. (a) The
conductance spectra in A-A(B) coupling for the incident electron
energy $E=0.0$, 0.1 and 0.5. (b) The case of Gaussian-type coupling
for $E=0$, 0.1 and 0.5.\label{c_r}}
\end{center}
\end{figure}
\begin{figure}
\begin{center}
\scalebox{0.5}{\includegraphics{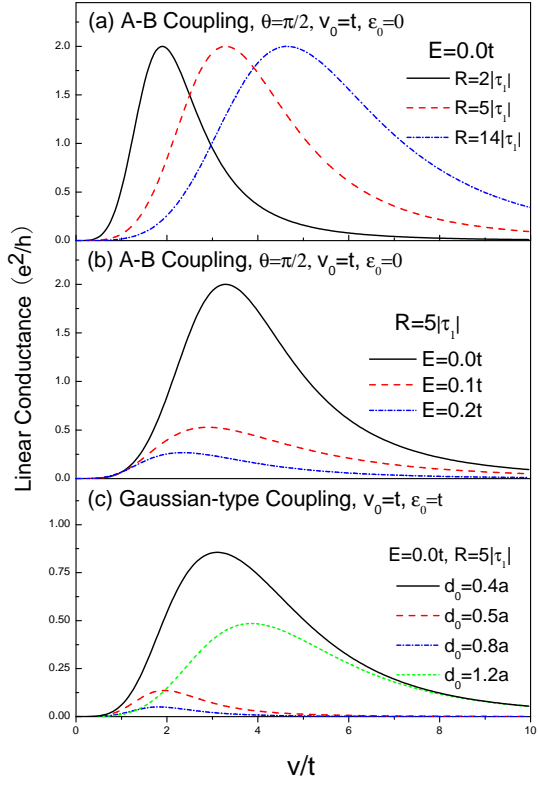}}\caption{(Color online)
The linear conductance as a function of $v$, the graphene-lead
coupling strength. (a) The case of A-B coupling, the incident
electron energy is fixed at $E=0$, but the probe interval varies
from $\bm{R}=2|\bm{\tau}_1|$, $|5\bm{\tau}_1|$ to $|14\bm{\tau}_1|$.
(b) A-B coupling configuration with the probe interval fixed at
$\bm{R}=|5\bm{\tau}_1|$, and the incident electron energy taking
several different values: E=0, 0.1 and 0.2. (c) Gaussian-type
coupling configuration with the probe interval and the incident
electron energy specified at E=0 and $\bm{R}=|5\bm{\tau}_1|$. The
decay factor in the Gaussian coupling function takes several
different values: $d_0=0.4$, 0.5, 0.8 and 1.2.\label{resonance}}
\end{center}
\end{figure}
\begin{figure}
\begin{center}
\scalebox{0.5}{\includegraphics{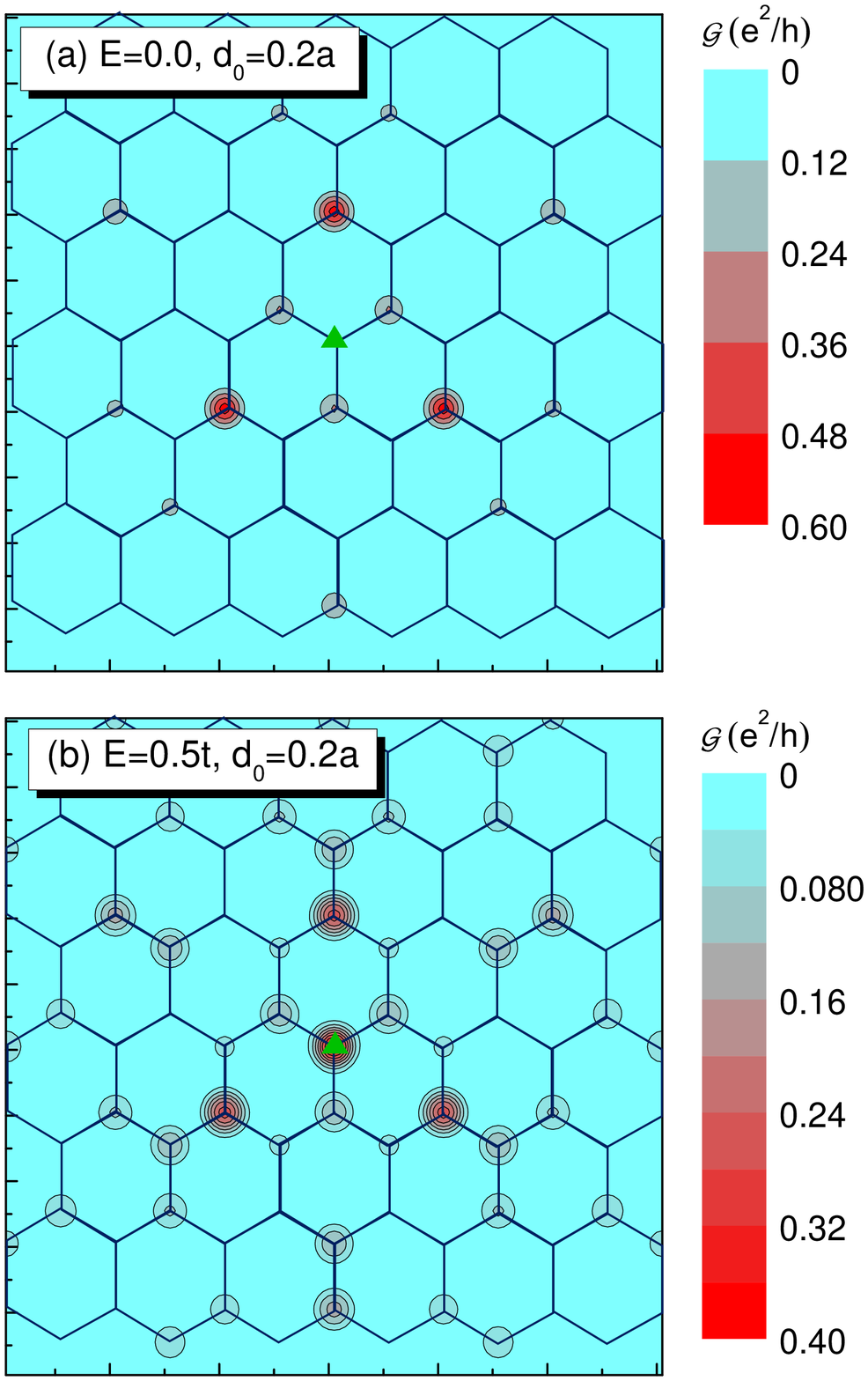}}\caption{(Color online)
The two-dimensional conductance pattern formed by fixing the first
probe at the origin labeled by a triangle, and shifting the second
probe around. (a) The case of $E=0$ and $d_{0}=0.2$. (b) The case of
$E=0.5$ and $d_{0}=0.2$. \label{patternd02}}
\end{center}
\end{figure}
\begin{figure}
\begin{center}
\scalebox{0.5}{\includegraphics{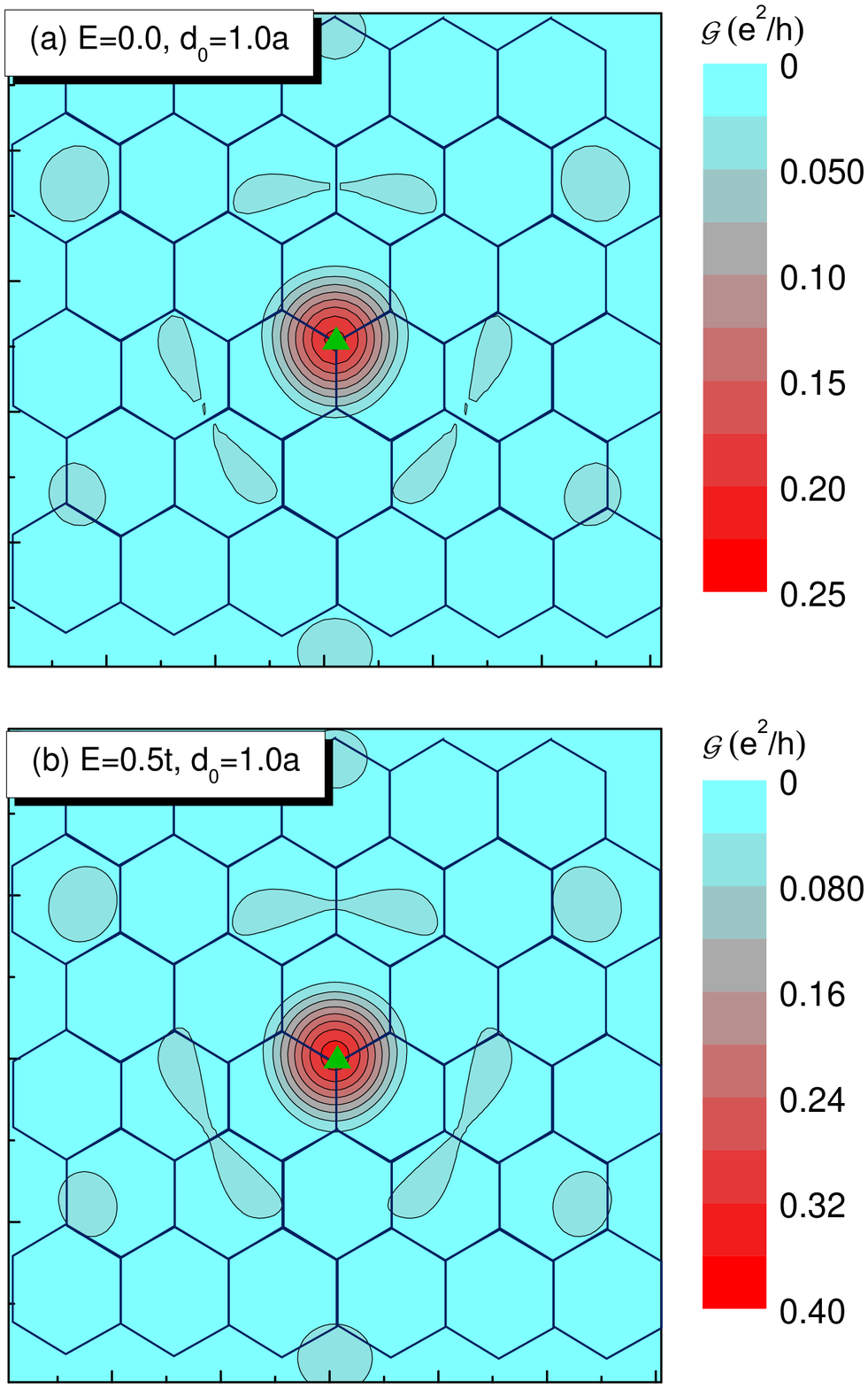}}\caption{(Color online)
The two-dimensional conductance pattern formed by fixing the first
probe at the origin labeled by a triangle, and shifting the second
probe around. (a) The case of $E=0$ and $d_{0}=1.0$. (b) The case of
$E=0.5$ and $d_{0}=1.0$.\label{patternd1}}
\end{center}
\end{figure}
\begin{figure}
\begin{center}
\scalebox{0.5}{\includegraphics{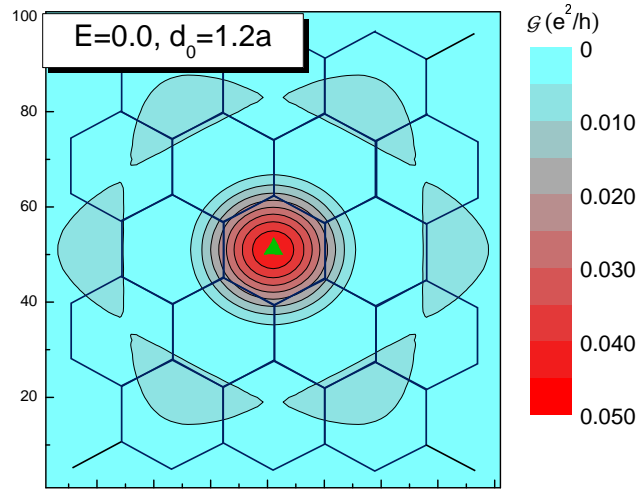}}\caption{(Color online)
The two-dimensional conductance pattern. As labeled by a triangle,
the first probe points at the center of a hexagon of graphene
lattice. $E=0$ and $d_{0}=1.2. $\label{patternring}}
\end{center}
\end{figure}

\end{document}